\documentclass[aps,prl, twocolumn, showpacs]{revtex4}
\usepackage{graphicx}
\usepackage{dcolumn}
\usepackage{bm}

\begin{document}
\preprint{preprint}
\title{Continuous loading of $^{1}$S$_{0}$ calcium atoms into an optical dipole trap}
\author{C. Y. Yang}
\author{P. Halder}
\author{O. Appel}
\author{D. Hansen}
\author{A. Hemmerich}
\affiliation{Institut f\"{u}r Laser-Physik, Universit\"{a}t Hamburg, 
Luruper Chaussee 149, 22761 Hamburg, Germany}
\date{\today}

\begin{abstract}
We demonstrate an efficient scheme for continuous trap loading based upon spatially selective optical pumping. We discuss the case of $^{1}$S$_{0}$ calcium atoms in an optical dipole trap (ODT), however, similar strategies should be applicable to a wide range of atomic species. Our starting point is a reservoir of moderately cold ($\approx 300 \, \mu$K) metastable $^{3}$P$_{2}$-atoms prepared by means of a magneto-optic trap (triplet-MOT). A focused 532 nm laser beam produces a strongly elongated optical potential for $^{1}$S$_{0}$-atoms with up to 350 $\mu$K well depth. A weak focused laser beam at 430 nm, carefully superimposed upon the ODT beam, selectively pumps the $^{3}$P$_{2}$-atoms inside the capture volume to the singlet state, where they are confined by the ODT. The triplet-MOT perpetually refills the capture volume with $^{3}$P$_{2}$-atoms thus providing a continuous stream of cold atoms into the ODT at a rate of $10^7\,$s$^{-1}$. Limited by evaporation loss, in 200 ms we typically load $5 \times 10^5$ atoms with an initial radial temperature of 85 $\mu$K. After terminating the loading we observe evaporation during 50 ms leaving us with $10^5$ atoms at radial temperatures close to 40 $\mu$K and a peak phase space density of $6.8 \times 10^{-5}$. We point out that a comparable scheme could be employed to load a dipole trap with $^{3}$P$_{0}$-atoms. 
\end{abstract}

\pacs{32.80.Pj, 34.50.-s, 82.20.Pm}

\maketitle
The unique spectroscopic features of two-electron systems and their usefulness for the fields of time metrology \cite{Did:04},  cold collision physics \cite{Wei:99, Bur:02} and quantum gases \cite{Ang:02} has led to extensive efforts to improve laser cooling and trapping techniques for alkaline earth (AE) atoms \cite{Kat:99, Bin:01, Cur:01, Gru:02, Han:03, Tak:03, Lof:04, Pol:05}. Calcium is a particularly interesting example, because, aside from its excellent performance in optical atomic clock scenarios \cite{Hel:03, Tai:06}, its singlet ground state (in contrast to the most abundant strontium isotope \cite{Mic:05}) has a large positive scattering length with favorable prospects for reaching quantum degeneracy \cite{Deg:03}. Optical trapping is a key technique in modern atomic physics, indispensible in numerous recent experiments with ultracold atoms and molecules \cite{Gor:80}. In particular, if magnetic trapping techniques fail to work, as in the singlet manifold of the AE group, optical dipole traps (ODTs) practically have no alternative. 

ODTs typically provide good compression, but suffer from limited trap depths of several hundred $\mu$K, owing to limitations in available laser powers. Thus, efficient loading of ODTs typically requires a magneto-optic trap (MOT) permitting sufficiently low temperatures well below 100 $\mu$K, as available for alkaline atoms. Although magneto-optical trapping of AE-like atoms in the ground state is in fact possible using their principal fluorescence lines, the attainable temperatures of several mK are too high for efficient direct loading of an ODT. In some cases, e.g. for strontium or ytterbium, additional cooling by means of intercombination lines connecting to their ground states \cite{Kat:99, Tak:03} have been used for ODT loading with large phase space densities. Unfortunately, this does not likewise apply to calcium, because of the low bandwidth (380 Hz) of its intercombination line, although low temperatures were reported \cite{Bin:01, Cur:01}.

\begin{figure}
\includegraphics[scale=0.4]{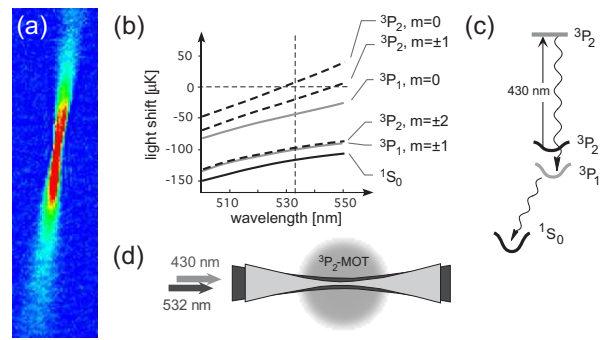}
\caption{\label{fig.1} (a) Absorption image of ODT. The line of sight encloses a small angle of $4^{\circ}$ with the ODT beam. (b) Light shifts of the relevant levels for 1 W and $\pi$-polarization. (c) Levels relevant for loading of ODT. (d) Geometry of the ODT and optical pumping beam.}
\end{figure}

In this article we demonstrate an efficient loading technique for an ODT of calcium atoms in the singlet ground state $^{1}$S$_{0}$, which has permitted us to obtain phase space densities around $7 \times 10^{-5}$. A conventional MOT for $^{1}$S$_{0}$-atoms (singlet-MOT), operating on the principle fluorescence line at 423 nm, is used to load a MOT for $\,[4s4p]^{3}$P$_{2}$-atoms (triplet-MOT) using infrared radiation at 1978 nm (for details see refs. \cite{Gru:02, Han:03}). A linearly polarized, focused laser beam ($w_0$ = 22.5 $\mu$m $1/e^2$ radius) at 532 nm with up to 3 W power produces a tight, strongly elongated, horizontally oriented light shift potential for $^{1}$S$_{0}$-atoms with up to 350 $\mu$K well depth. Negative light shifts also arise for the $^{3}$P$_{2}$ and the $^{3}$P$_{1}$ levels (cf. Fig.\ 1(b)). Atoms are loaded into the ODT by optical pumping using a weak violet laser beam at 430 nm with a $1/e^2$ beam radius $w_1$ = 17 $\mu$m. This beam resonantly excites the $\,[4s4p]^{3}$P$_{2}$-atoms to $\,[4p^2]^{3}$P$_{2}$ from where they may decay to the singlet ground state via $\,[4s4p]^{3}$P$_{1}$ (cf. Fig.\ 1(c)). As sketched in Fig.\ 1(d), the optical pumping beam is shaped to match the ODT potential, such that only $^{3}$P$_{2}$-atoms located within the capture volume are pumped to the light shift potential of the $\,[4s4p]^{3}$P$_{1}$ level and further decay to the $^{1}$S$_{0}$ ODT within 0.42 ms. The capture volume perpetually refills with $^{3}$P$_{2}$-atoms from the triplet-MOT, i.e., a continuous stream of cold atoms is transferred to the ODT. The population of the ODT exponentially approaches a steady state determined by the balance between loading and evaporation loss, long before the $^{3}$P$_{2}$-population of the triplet-MOT of several $10^8$ atoms is exhausted. We obtain an excellent loading efficiency with capture rates of up to $10^7 s^{-1}$ and typically $5 \times 10^5$ trapped atoms in steady state, despite the unfavorable matching of the ODT and MOT volumes, the relatively low trap depth and the large cross-section for inelastic collisions between $\,[4s4p]^{3}$P$_{2}$-atoms \cite{Han:06}. A related technique has recently been used for loading an optical lattice of strontium atoms \cite{Bru:06}, however, at a hundredfold lower capture rate of $10^5$ s$^{-1}$. 

\begin{figure}
\includegraphics[scale=0.40]{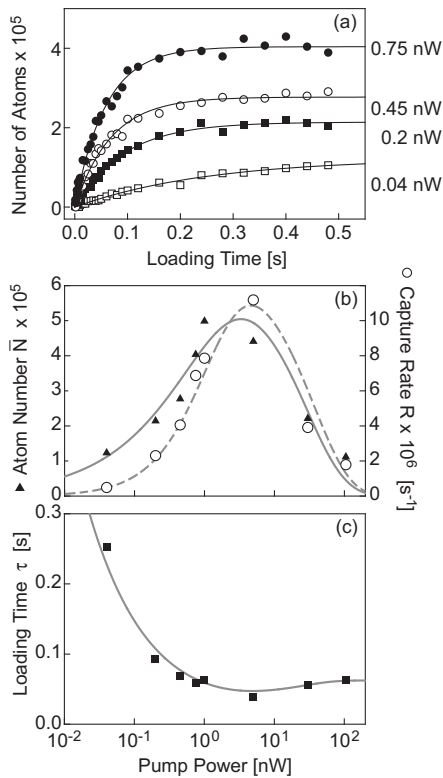}
\caption{\label{fig.2} In (a) trap loading measurements are shown for different values of the optical pumping power indicated on the right margin. The solid lines are exponential fits. Below, the steady-state number of atoms loaded into the dipole trap (b) and the 1/e loading time (c) are plotted versus the power of the optical pumping beam. The grey lines are derived with the model described in the text.}
\end{figure}

We have studied the exponential loading dynamics at varying optical pumping powers as follows. Initially, the triplet-MOT is loaded for 1 s by the singlet-MOT. Subsequently, the singlet-MOT is extinguished, the ODT beam and the pumping beam are enabled. The negative frequency detuning of the triplet-MOT beams is increased to compensate for the negative light-shift of the triplet-MOT transition within the capture volume, at the cost of an undesired increase of the triplet-MOT temperature outside the capture volume to values around 300 $\mu$K. After a variable time loading is terminated and an absorption image is taken, which lets us derive the instantaneous number of atoms in the ODT. Typical examples of the observed exponential loading curves are plotted in Fig.\ 2(a). Such curves let us determine the number of atoms $\bar N$ loaded into the ODT in steady state (black triangles in Fig.\ 2(b)), the capture rate $R$ (open circles in Fig.\ 2(b)), obtained as the initial slope of the exponential fits in Fig.\ 2(a), and the 1/e loading time $\tau$ (black squares in Fig.\ 2(c)). For very low powers $R$ increases with power yielding an increase of $\bar N$ and a decrease of $\tau$. The increase of $R$ arises because below saturation the optical pumping rate is proportional to the light intensity. The consequential increase of $\bar N$ yields an increase of the density in the ODT and thus an increase of evaporative loss, which in turn yields a decrease of $\tau$. When saturation of the pumping transition is approached, the effective pumping volume begins to exceed the trap volume, i.e., many atoms are pumped outside the trap volume of the ODT and cannot be captured. Thus, $R$ decreases again for high pump powers, which should be accompanied by an increase of $\tau$ due to reduced evaporation. The observations in Fig.\ 2(c) indicate that for large pump powers an additional loss mechanism sets in, which keeps the value of $\tau$ smaller than expected from evaporation of the relatively small ODT population. We speculate that the increased atom population pumped outside the ODT volume acts as an additional hot background introducing extra losses. 

The preceeding considerations can be rendered more precisely by extending a common model applied in numerous previous works  \cite{Din:99, Gru:01}. The instantaneous number of atoms $N$ evolves according to
\begin{equation}
\label{trap_decay} 
\frac{d}{dt} N   =  R(P) - (\gamma + \gamma_B(P))\,N - \gamma_{ev}(T)\,N^2  \,\,, 
\end{equation}
where $\gamma$ is the loss rate due to collisions with fast background gas atoms, $\gamma_B(P)$ is the loss rate due to collisions with untrapped $^{1}$S$_{0}$-atoms optically pumped outside the trap volume, and $\gamma_{\rm{ev}}(T)$ is a temperature-dependent two-body loss parameter accounting for evaporative loss via elastic binary collisions between trapped singlet atoms. The dependences of the capture rate $R(P)$ and the loss rate $\gamma_B(P)$ on the pump power $P$ are calculated by integrating the optical pumping rates for atoms in the triplet-MOT inside and outside the ODT volume. In order to calculate $\gamma_{\rm{ev}}(T)$, an energy-independent elastic scattering cross-section $\sigma_{0}$ is assumed. The fraction of collisions yielding evaporation is obtained according to a model of detailed balance \cite{Ket:96} based upon a harmonic approximation of the ODT potential. Assuming a constant temperature, Eq.\ (\ref{trap_decay}) is solved, in order to find the 1/e loading time $\tau$ and the steady state particle number $\bar N$ (see e.g. \cite{Gru:01}). 

To compare our model with the observations, three fit-parameters are employed: the absolute scales of $R(P)$ and $\gamma_B(P)$ and the constant elastic scattering cross-section $\sigma_0$. First, we adjust the absolute scale of $R(P)$, in order to optimize the agreement with the observed capture rates (open circles in Fig.\ 2(b)). This yields the dashed line in Fig.\ 2(b). In a second step the overall scale of $\gamma_B(P)$ and the scattering cross-section $\sigma_{0}$ are adjusted to obtain the solid line in Fig.\ 2(c). In this graph $\sigma_{0} = 2.9 \times 10^{-16}$ m$^2$, which corresponds to the unitarity limit evaluated for a temperature of 175 $\mu$K. Recent calculations and measurements have set boundaries for the s-wave scattering length $a_{\rm{scat}}$ between 340 $a_0$ and 800 $a_0$ ($a_0$ = Bohr-radius) \cite{Deg:03}. A detailed model connecting $a_{\rm{scat}}$ to the scattering cross-section is not available for $^{1}$S$_{0}$ calcium atoms, however, the size of $\sigma_{0}$ appears compatible with the measured range of $a_{\rm{scat}}$. Finally, the calculated steady state number of atoms $\bar N(P)$ is scaled by a constant factor 0.53 to match with the observed data. We believe that this reflects a systematic error in the absolute calibration of our absorption imaging system used to measure $\bar N$. As fixed parameters of the model we use a well depth of 350 $\mu$K, exceeding the value obtained from measurements of the power and the focus of the ODT beam by $15 \%$, and a temperature $T_0 \equiv  (T_{\rm{ax,0}} + 2 \, T_{\rm{rad,0}})/3 = 157 \, \mu$K resulting from the initial axial and radial temperatures $T_{\rm{ax,0}} = 300 \, \mu$K and $T_{\rm{rad,0}} = 85 \, \mu$K. The used radial temperature results from measurements described below. Its value well below that of the triplet-MOT of about 300 $\mu$K reflects the fact that atoms with small radial velocities remain longer in the pumping beam and are thus more efficiently loaded. Due to the extreme aspect ratio of the ODT this does not equally apply to the axial direction and thus $T_{\rm{ax,0}}$ has been assumed to be equal to the triplet-MOT temperature. The good agreement between the calculated functions $R(P)$, $\tau(P)$ and $\bar N(P)$ and the observations shows that the model despite its simplifications identifies the relevant physical mechanisms. It provides the useful insights that loading is mainly limited by evaporation and that different axial and radial temperatures are produced.

\begin{figure}
\includegraphics[scale=0.4]{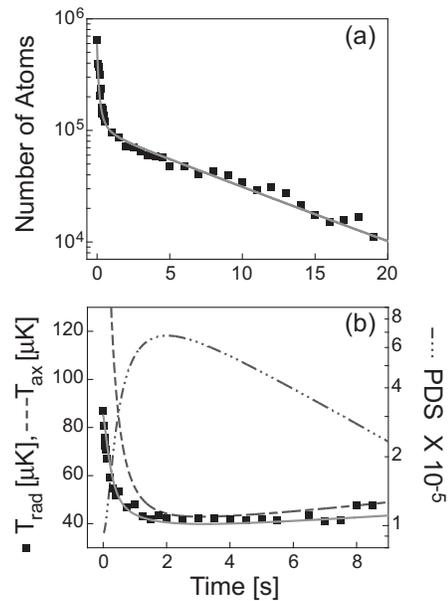}
\caption{\label{fig.3} The observed number of atoms (filled squares in (a)) and the radial temperature (filled squares in (b)) are plotted versus time after loading is terminated. The grey lines are calculated with the model described in the text.}
\end{figure}

We have also observed the time evolution of the particle number and the radial temperature of the ODT after loading is terminated. In these measurements the atoms are kept in the ODT for a variable time and subsequently a time-of-flight (TOF) method is applied. In the TOF procedure the ODT potential is switched off, the atoms are given the chance to expand ballistically for some period finished by taking an absorption image. A series of TOF measurements for a fixed hold-up time of the atoms in the ODT lets us derive the instantaneous radial temperatures and particle numbers. In Fig.\ 3 we have recorded the trap loss (filled squares in (a)) and the evolution of the radial temperature (filled squares in (b)) after loading is terminated. In (a) one recognizes a fast loss of atoms during the first 50 ms, which results from evaporation accompanied by a significant reduction of the initial radial temperature $T_{\rm{rad,0}} = 85 \, \mu$K to values close to 40 $\mu$K (cf. (b)). 

In order to calculate the ODT dynamics after termination of loading, we may complement Eq.\ (\ref{trap_decay}) by two further equations describing the time evolution of the radial ($T_{\rm{rad}}$) and axial ($T_{\rm{ax}}$) temperatures
\begin{eqnarray}
\label{energy_budget} 
\dot{T}_{\rm{ax}} & = & - \frac{1}{3} \gamma_{\rm{ev}}\, N\,(\frac{T_{\rm{ev}}}{T}-1)\, T_{\rm{ax}} -\frac{2}{3} \gamma_{\rm{rel}}\,\Delta T+\frac{4}{3} \gamma_{\rm{sc}}\, T_{\rm{rec}}\,\, ,
\nonumber \\  \\   \nonumber
\dot{T}_{\rm{rad}} & = & - \frac{1}{3} \gamma_{\rm{ev}}\, N\,(\frac{T_{\rm{ev}}}{T}-1)\, T_{\rm{rad}}  +\frac{1}{3} \gamma_{\rm{rel}}\,\Delta T+\frac{2}{3} \gamma_{\rm{sc}}\, T_{\rm{rec}}\,\, .
\end{eqnarray}
Here, $T \equiv (T_{\rm{ax}} + 2 \, T_{\rm{rad}})/3$, $\Delta T \equiv T_{\rm{ax}}-T_{\rm{rad}}$, $k_B T_{\rm{ev}}$ is the energy removed per evaporated particle, $k_B T_{\rm{rec}}$ is the photon recoil energy, $\gamma_{\rm{sc}}$ is the photon scattering rate of the ODT (9 s$^{-1}$), and $\gamma_{\rm{rel}}$ is the cross-dimensional relaxation (CR) rate. In Eq.\ (\ref{energy_budget}) exponential thermalisation of $T_{\rm{rad}}$ and $T_{\rm{ax}}$ is assumed, proven to be a suitable approximation in numerous previous CR-studies \cite{Mon:93,Hop:00,Han:06}. The CR-rate is calculated from the energy-independent scattering cross-section $\sigma_0$ with the help of Ref.\ \cite{Kav:00}. The values of $\gamma_{\rm{ev}}$ and $T_{\rm{ev}}$ are obtained from the detailed balance model \cite{Ket:96} used in the context of Eq.\ (\ref{trap_decay}). Using the same initial axial and radial temperatures ($T_{\rm{ax,0}} = 300 \, \mu$K, $T_{\rm{rad,0}} = 85\, \mu$K) and the value of $\sigma_0$ as in the calculated graphs of Fig.\ 2(b,c), the system of  Eq.\ (\ref{trap_decay}) with $R(P) = 0$ and Eq.\ (\ref{energy_budget}) is solved numerically. As done in Fig.\ 2(b) for $\bar N$, we finally scale the calculated values for $N$ by the calibration factor 0.53. The results for $N$ and $T_{\rm{rad}}$ (solid lines in (a) and (b)) are shown in Fig.\ 3. Our calculations also provide the evolution of the axial temperature (dashed line in (b)), which cannot be observed in our experiment. This lets us obtain the phase space density $PSD \equiv N \hbar^3 \Omega_{\rm{ax}}\Omega_{\rm{rad}}^2/(k_B T)^3$ (dashed-dotted line in (b)), where $\Omega_{\rm{ax}} = 2 \pi \times 20.3$ Hz and $\Omega_{\rm{rad}}= 2 \pi \times 3.8$ kHz are the calculated axial and radial harmonic frequencies of the ODT.

\begin{figure}
\includegraphics[scale=0.40]{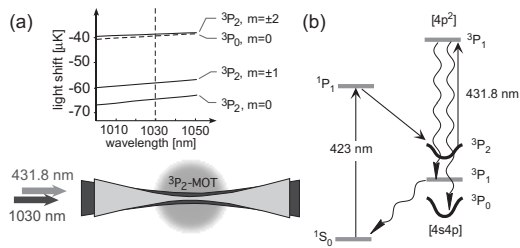}
\caption{\label{fig.4} Continuous loading scheme for an ODT of $^{3}$P$_{0}$-atoms. (a) Light shifts and sketch of beam configuration. (b) Relevant electronic levels.}
\end{figure}

Finally, Fig.\ 4 indicates, how an ODT could be continuously loaded with $^{3}$P$_{0}$-atoms. This state could play a central role in future optical clocks exceeding $10^{-17}$ relative precision \cite{Tai:06}. A dipole trap using 1030 nm light (readily available with sufficient power) produces nearly equal light shifts for $^{3}$P$_{2},m=\pm2$ and $^{3}$P$_{0},m=0$ atoms (cf. (a)). Optical pumping between these states is achieved by means of 431.8 nm radiation coupling $\,[4s4p]^{3}$P$_{2}$ to $\,[4p^2]^{3}$P$_{1}$ (cf. (b)), which (apart from ppm contributions) decays either to $\,[4s4p]^{3}$P$_{1}$ or $\,[4s4p]^{3}$P$_{0}$ with a branching ratio of 44\% to 56\%. Both the triplet-MOT and the singlet-MOT are kept active during the entire loading. Atoms transferred to $\,[4s4p]^{3}$P$_{1}$ subsequently decay to $^{1}$S$_{0}$ and are thus readily transferred back to the cold $^{3}$P$_{2}$-reservoir.

In summary, we have demonstrated continuous loading of $^{1}$S$_{0}$ calcium atoms into an optical dipole trap operating at 532 nm with capture rates of up to $10^7 s^{-1}$. Metastable $^{3}$P$_{2}$-atoms precooled by a magneto-optic trap are loaded into the dipole trap by spatially selective optical pumping using a weak focused laser beam at 430 nm spatially matched to the trapping beam. We typically load $5 \times 10^5$ atoms with initial temperatures around 300 $\mu$K in axial and 85 $\mu$K in radial directions of the dipole trap. During 50 ms after terminating the loading we observe evaporation and cross-dimensional relaxation leaving us with $10^5$ atoms at 41 $\mu$K and a phase space density of $6.8 \times 10^{-5}$. A simple model shows that evaporation is the dominant loss mechanism. We have pointed out how continuous loading can be applied to the case of a dipole trap with $^{3}$P$_{0}$-atoms, which could be a useful tool for time metrology. Similar schemes should be applicable to a wide range of atomic species. 

\begin{acknowledgments}
This work has been supported by DFG (\textit{priority program SPP 1116, He2334/9-1}). C. Y. Yang acknowledges a fellowship by DAAD.
\end{acknowledgments}

\end{document}